\documentclass[twocolumn,pra,amssymb,]{revtex4}

\usepackage{graphicx}
\usepackage{dcolumn}
\usepackage{epsfig}
\usepackage{color}

\def\etal{{\it et al.}}                                        %

\begin{document}

\preprint{}

%
%

\title{Electron loss of fast heavy projectiles in collision
       with neutral targets}

%
%

\author{V.~I.~Matveev}
\affiliation{Physics Department, Pomor State University, 163002
             Arkhangelsk, Russia}

\author{S.~V.~Ryabchenko}
\affiliation{Physics Department, Pomor State University, 163002
             Arkhangelsk, Russia}

\author{D.~U.~Matrasulov}
\affiliation{Heat Physics Department of the
             Uzbek Academy of Sciences,
             100135 Tashkent, Uzbekistan}

\author{Kh.~Yu.~Rakhimov}
\affiliation{Heat Physics Department of the
             Uzbek Academy of Sciences,
             100135 Tashkent, Uzbekistan}

\author{S.~Fritzsche}
\affiliation{Frankfurt Institute of Advanced Studies (FIAS)
             D-60438 Frankfurt am Main, Germany}
\affiliation{Gesellschaft f\"ur Schwerionenforschung (GSI),
             D-64291 Darmstadt, Germany}

\author{Th.~St\"ohlker}
\affiliation{Gesellschaft f\"ur Schwerionenforschung (GSI),
             D-64291 Darmstadt, Germany}
\affiliation{Physikalisches Institut der
             Universit\"{a}t Heidelberg,
             D-69120 Heidelberg, Germany}

\date{\today \\[0.3cm]}

%
%

%
%
%
\begin{abstract}
The multiple electron loss of heavy projectiles in fast ion-atom collisions
has been studied in the framework of the sudden perturbation approximation.
Especially, a model is developed to calculate the
cross sections for the loss of any number of electrons from the projectile
ion, including the ionization of a single electron and up to the complete
stripping of the projectile. For a given collision system, that is specified
by the (type and charge state of the) projectile and target as well as the
collision energy, in fact, the experimental cross sections for just three
final states of the projectile are required by this model in
order to predict the loss of any number, $N$, of electrons for the same
collision system, or for any similar system that differs only in the energy
or the initial charge state of the projectile ion. The model is simple
and can be utilized for both, the projectile and target ionization, and
without that large computer resources are requested. Detailed computation
have been carried out for the multiple electron loss of Xe$^{18+}$ and
U$^{6+,\, 10+,\, 28+}$ projectiles in collision with neutral Ar and Ne gas
targets.
\end{abstract}

 \maketitle

\section{Introduction}

In recent years, the loss of electrons from heavy projectiles in
fast collisions with neutral target atoms has attracted a great deal of
interest because of its importance for the design of new heavy-ion
accelerators and storage rings~\cite{fair1,fair2,fair3}. In such rings,
any ionization or capture of electrons by the projectiles typically leads
to a loss of the ions from the beam, thus limiting the intensities
that are to be obtained eventually. For the International Facility of
Antiproton and Ion Research (FAIR) at the GSI in Darmstadt, for example,
reliable predictions are required especially for the multiple electron
loss of fast uranium ions in collision with neutral rest gas atoms
in order to estimate the beam intensities that can be
achieved and manipulated at the SIS and the subsequent
experimental storage rings~\cite{Armen,gsirep}. Indeed, many
atomic and nuclear experiments that are planned for this facility will
depend on the availability of having intensive beams with projectile
energies of several ten MeV/u to a few GeV/u, and on the excellent
control of all the beam parameters~\cite{Stoehlker:04,Fritzsche:05}.

For these reasons, several experiments have been performed in the
past in order to measure and analyze the multiple loss of
electrons for different projectiles and various rest gas and foil
targets~\cite{Berg,Ullrich,DuBois,Watson}. In particular, a series
of measurements were carried out for highly-charged uranium ions
in collision with neon, argon and krypton targets in order to make
available a systematic set of total ionization cross
sections~\cite{Berg,Ullrich}. Most of these measurements were
found in a reasonable-to-good agreement with Classical Trajectory
Monte Carlo (CTMC) computations. Moreover, multiple and total
(absolute) cross section have been measured also for 1.4~MeV/u
uranium ions in collision with Ne, Ar and N$_2$
targets~\cite{DuBois}. In these experiments, it was shown
especially that, for 1.4~MeV/u U$^{4,6,10+}$ ions, the cross
sections for a simultaneous loss of electrons first decrease
rather slowly in magnitude if only up to about 10 electrons are
stripped from the projectile, while they fall off much more
rapidly for all higher final charge states of the projectiles. In
addition, Watson \etal{}~\cite{Watson} have measured recently the
electron loss cross sections for 6~MeV/u Xe$^{18+}$ projectiles in
collisions with (noble) He, Ne, Ar, Kr, and Xe gas targets.

From the viewpoint of atomic theory, of course, the main
difficulty in predicting the (simultaneous) loss of several
electrons in collisions of high-Z projectiles with light rest gas
atoms or target material arises from the break-down of
perturbation theory if the charge (state) of the final ions is
increased successively. For U$^{10+}$ ions, for example, a
perturbation approach breaks down already for all collision
energies smaller than about E=2.5~MeV/u. For the ion beams at the
FAIR facility in Darmstadt, therefore, the Born approximation
cannot be applied since its basic condition, $Z/v \ll 1$, is no
longer fulfilled for high-Z projectiles even for rather moderate
velocities. For this reason, it is necessary to explore further
models that can be utilized beyond the Born approximation, and to
estimate the total and differential cross sections on more
reliable grounds.

Until now, several computational models have been developed to
calculate the multiple electron loss of fast projectiles in
collision with different target materials: Apart from the
classical trajectory Monte Carlo method~\cite{Olson} and the
computer code \textsc{Loss}, implemented by Shevelko and
coworkers~\cite{Shevelko}, the (so-called) sudden perturbation
approximation has been worked out
recently~\cite{VI05,Matv2,Matvgsi} that enables one in an
efficient way to estimate the inelastic transition amplitudes and
cross sections in collisions of fast projectiles with different
target atoms. In this previous work~\cite{Matv2}, however, only
the active electrons were taken into account in the theoretical
treatment, i.e.\ those which are ionized in course of the
collision, ignoring the remaining charge density at the
projectile. In the present work, we now release this restriction
and account for all of the electrons in order to re-analyze the
energy-dependence of different electron loss cross sections which
are known from experiment. Moreover, we have extended the sudden
approximation to take into account the multiple electron loss from
both, either the projectile or the target ions, if some measured
cross sections are known already for just three final states of
the ions from prior experiments. In fact, the model described
below is simple and can be utilized in order to predict the loss
of any number, $N$, of electrons for either the same collision
system (as specified by the type and incoming charge states of the
projectile and target atoms as well as their collision energy), or
for any similar system that only differs in the energy, the
initial charge state of the projectiles, or in the choice of the
target atoms.

The paper is organized as follows: In the next section, we first
recall the basic ideas of the \textit{sudden} (perturbation)
approximation and how this method can be utilized in order to
calculate the cross sections for a multiple electron loss from fast
projectiles. During the last years, this method was developed
especially for studying the multiple ionization processes of
either the target and/or the projectiles and for the case, that
the (orbital) motion of the bound electrons can be neglected when
compared with the relative velocity of the collisions
partners~\cite{Matv2}. Apart from the transition amplitude for the
projectile electron loss, this section also formulates a recipe how
the cross section for a multiple loss of electrons can be predicted, if
analogue cross sections are known already for the same collision partners
but for a different charge state or energy of the projectiles ions.
In Section~III, this method is then utilized to calculate the multiple
electron loss cross sections for fast uranium and xenon projectiles.
Results are shown for different collision energies, several initial
charges states of the projectile ions and for Ar and Ne targets,
respectively. Finally, a few conclusions are drawn in section~IV.

\section{Theoretical background}

Originally, the sudden perturbation approximation was developed to
efficiently describe the multiple electron loss from target atoms
in fast ion-atom collisions~\cite{Matv0,Matv2}. This approximation
is based on the assumption that a (not necessarely weak)
time-dependent perturbation acts on a collision system for a time
which is \textit{much shorter} than the period of all the
electrons to be ionized in course of the collision. Making use of
this assumption, the transition amplitudes for the projectile
ionization/excitation can be evaluated without that the
time-dependent Schr\"odinger or Dirac equation would need to be
solved explicitly for the ion-atom collision. In a first
application, the sudden perturbation approximation was employed to
the target excitation and ionization~\cite{Matv0}, including the
simultaneous stripping of several electrons, but has been utilized
more recently also in order to calculate the electron loss of
heavy projectiles in fast collisions with neutral
targets~\cite{VI05,Matv2,Matvgsi,Matv4}. In the present work, we
aim to extent this formalism to the case that several or even all
electrons are ionized simultaneously from the projectile and that
all the electrons are taken into account in terms of their
mean-field.

To describe quantitatively a multiple ionization of $N$ electrons
from fast projectiles, let us consider a collision system in which
a projectile with charge $Z_p$ and initially $N_p \:(\ge N)$
electrons collides with a neutral target with charge $Z_a$ and
$N_a$ electrons. In the following, moreover, let us assign to the
electrons of the projectile and target the coordinates ${\bf r}_p$
($p=1,2,...,N_p$) and ${\bf r}_a$ ($a=1,2,...,N_a$), and which are
defined in the projectile and target frame, respectively. If the
projectile nucleus moves with constant (relative) ${\bf v}$ and
impact parameter (closest approach) ${\bf b}$ along a
semi-classical trajectory, the interaction potential between the
target and projectile can then be written as
\begin{eqnarray}
\label{eqq1}
   V & = & -\sum_{a=1}^{N_a} \: \frac{Z_p}{|{\bf R}(t)+{\bf r}_a|}
      \:-\: \sum_{p=1}^{N_p} \: \frac{Z_a}{|{\bf R}(t)+{\bf r}_p|}
   \nonumber \\[0.2cm]
   &  & \hspace*{0.5cm}
   + \: \sum_{p,a=1}^{N_p,\, N_a} \:
        \frac{1}{|{\bf R}(t)+{\bf r}_p-{\bf r}_a|} \, ,
\end{eqnarray}
if ${\bf R}(t) = {\bf b} + {\bf v}t$ denotes the (time-dependent)
distance between the two nuclei, and where we make use of atomic
units (if not stated otherwise). Obviously, this interaction
potential consists of three terms where the first one describes the
interaction of the target electrons with the nucleus of the projectile,
the second ---\textit{vice versa}--- the interaction between the projectile
electrons and the target nucleus, and the third term finally the
pairwise repulsion between the electrons with one belonging to the
projectile and the other to the target, respectively.

In order to make use of the sudden approximation, we shall assume
moreover that the (interaction) collision time $t_c$ between the
projectile and the target is much shorter than the period $t_s$ of
the most inner (and fastest) electron to be ionized from the
projectile, i.e.\
\begin{eqnarray}
\label{tau1}
   t_c \ll t_s \, .
\end{eqnarray}
Since the collision time is roughly given by $t_c \,\approx\,
a/v$, with $a \approx 1$ (in atomic units) being the
characteristic size of the collision partners and $v=|{\bf v}|$
their relative velocity, and since the period of all orbital is
$t_s\lesssim 1$ for all (positive) ions, the condition
(\ref{tau1}) becomes $t_c \approx 1/v \ll 1$ or, equivalently,
$$  v \gg 1.  $$
This means that, as mentioned before, the sudden approximation
can be safely applied only if the relative projectile-target
velocity is (much) larger than the velocity of the electrons to be
ionized  during the collision.

For a large enough distance of the two collision partners, of
course, wave functions can be assigned independently to both, the
projectile and target. Below, let us suppose to have a complete
set of wave functions $\psi_k  = \psi_k(\{{\bf r}_p\})\equiv
\psi_k({{\bf r}}_{1},
 \ldots,{{\bf r}}_{N_p})$ for the projectile electrons, and
a certain number of wave functions (state vectors) $\varphi_n
= \varphi_n(\{{\bf r}_a\})\equiv \varphi_n({{\bf r}}_{1},
 \ldots,{{\bf r}}_{N_a})$ for the target, including with $\psi_0$,
respectively, $\phi_0$ the corresponding (and undisturbed) ground
states in both cases. With this notation, the probability for any
excitation or ionization of the target from state $\varphi_0$ to
$\varphi_n$ and the projectile from $\psi_0$ to $\psi_{k}$ is
given in perturbation theory by
\begin{eqnarray}
w_{0\to{k}}^{0\to n}=
|\langle\varphi_n\psi_k|\exp\left(-i\int_{-\infty}^{+\infty}
Vdt\right)|\psi_{0}\varphi_0\rangle|^2.
\end{eqnarray}
Since, in the following, we are not further interested in the
final state of the target, we then obtain the probability for a
transition of the projectile from the ground state $\psi_0$ to
some excited and/or ionized state $\psi_{k}$ by a summation over
all final states
$$
W_{0\to k} \;=\; \sum_{n} \: w_{\, 0\to{k}}^{\, 0\to n} \, .
$$

As shown in further detail in Ref.~\cite{Matv2}, this
excitation-ionization probability of the projectile can be
expressed as function of the impact parameter ${\bf b}$ as
\begin{eqnarray}
\label{8ma}
   W_{0\to k}=\langle\varphi_0|\left|\langle\psi_{k}|
   \exp\left(-i\int_{-\infty}^{+\infty} U_adt\right)
   |\psi_{0}\rangle\right|^2|\varphi_0\rangle \, ,
\end{eqnarray}
if we assume that the sudden perturbation approximation is valid
and if the interaction potential
\begin{eqnarray}
\label{9a}
   U_a =  -\sum_{p=1}^{N_p} \frac{Z_a}{|{\bf R}(t)+{\bf r}_p|}
   + \sum_{p,a=1}^{N_p,N_a}\frac{1}{|{\bf R}(t)+{\bf r}_p-{\bf r}_a|}\, ,
\end{eqnarray}
with ${\bf R}(t) = {\bf b} + {\bf v}t$, now includes only the
interaction between the projectile electrons and the target
nucleus as well as the interelectronic repulsion.

Despite of its approximate validity, the direct use of
formula~(\ref{8ma}) is hardly feasible, especially if there are
many electrons involved in the target and projectile. Therefore,
in order to further simplify this expression, we shall assume in
the following that the positions of the projectile electrons
\textit{do not} change with regard to the target nucleus during
the time of the collision. In this case, it becomes possible to
average over the interaction potential between the target nucleus
and the projectile electrons and over the initial state of the
target electrons. Here, moreover, we shall suppose also that the
state of the target electrons are well described by means of the
one-electron orbitals as obtained from the Dirac-Hartree-Fock
model, forming a mean-field target density. Under these assumption
a much simpler formula were derived earlier~\cite{Matv2}
\begin{eqnarray}
\label{8m}
   W_{0\to k} & = &
   \left| \langle \psi_{k}|
                  \exp\left(-i\int_{-\infty}^{+\infty} \overline{U}_adt\right)|
                  \psi_{0}
   \rangle\right|^2 \, ,
\end{eqnarray}
in which the averaged potential, $\overline{U}_a$, is given by
\begin{eqnarray}
\label{9aa}
   \overline{U}_a & = &
   \langle\varphi_0(\{{\bf r}_a\})|U_a|\varphi_0(\{{\bf r}_a\})\rangle
   \nonumber  \\[0.2cm]
   & = & -\sum_{p=1}^{N_p} \: \frac{Z_a}{|{\bf R}(t)-{\bf r}_p|} \:
   \sum_{i=1}^{3} \, A_ie^{-\alpha_i|{\bf R}(t)-{\bf r}_p|} \, ,
\end{eqnarray}
and where the second line was obtained by applying the
parameterized Dirac-Hartree-Fock-Slater ground state
wavefunctions~\cite{salvat}. Let us note here that Eq.~(\ref{8m})
formally coincides also with Glauber approximation in which the
energy differences between the projectile states are neglected and
the target was supposed to be frozen in its initial state (see
Ref.~\cite{glauber} for further details). In Eq.~(\ref{9aa}),
moreover, the averaged potential $\overline{U}_a$ does no longer
depend on the coordinates $\{{\bf r}_a\}$ of the target electrons
but only on some tabulated constants $A_i$ and $\alpha_i$ as
listed for different atoms in Ref.~\cite{salvat}. In fact, taking
the average in Eq.~(\ref{9aa}) implies that the target has a
ground-state charge density $-\rho_a(r)$ of the form~\cite{Matv2}
\begin{eqnarray}
\label{9bb}
   \rho_a({\bf r}) & = &
   \frac{Z_a}{4\pi|{\bf r}|} \:
   \sum_{i=1}^{3} \: A_i{\alpha_i}^2e^{-\alpha_i|{\bf r}|} \, ,
\end{eqnarray}
as parameterized by means of the Dirac-Hartree-Fock-Slater model.

Using Eq.~(\ref{9bb}), it can be shown~\cite{Matv0} moreover that
the integral in Eq.~(\ref{8m}) can be written also as a sum of
`eikonal phases' of the projectile electrons
\begin{eqnarray}
   \int_{-\infty}^{+\infty}\overline{U}_adt=
   \sum_{p=1}^{N_p}\chi({\bf b},{\bf r}_p)\, ,
   \label{8mha}
\end{eqnarray}
with the functions $\chi({\bf b},{\bf r}_p)$ given by
\begin{eqnarray}
\label{21}
   \chi({\bf b},{\bf r}_p) & = &
   -\frac{2Z_a}{v} \: \sum_{i=1}^{3} \:
   A_i \, K_0(\alpha_i| {\bf b}-{\bf s}_p|) \, .
\end{eqnarray}
In this formula, ${\bf s}_p$ denotes the projection of ${\bf r}_p$
onto the plane that is perpendicular to the velocity (vector) of
the projectile and $K_0$ is the lowest-order McDonald function.
The (total) cross section for the ionization of the projectile,
and averaged over all states of the target electrons, is then
obtained by integrating the probability in Eq.~(\ref{8m}) over all
impact parameters
\begin{eqnarray}
\label{sigm}
   \sigma & = &  \int d^2\,{\bf b} \; W(b)
     \; \equiv\; \int d^2\,{\bf b} \;
                 \sum_{\quad[k]}\hspace*{-0.55cm}\int \: W_{0\to{k}} \, ,
   \label{80ha}
\end{eqnarray}
and where the restricted summation (integration) over $[k]$ runs
over those final state of the projectile electrons, where a given
number of electrons, $N$, have left the projectile. In
Eq.~(\ref{80ha}), this means that we are not interested in the
momenta of the outgoing electron but only in the dependence of the
cross sections on the charge $Z_p$ and the (relative) velocity $v$
for just the simultaneous loss of $N$ electrons from the
projectile ions.

Using Eqs.~(\ref{8m})-(\ref{sigm}), the cross section for a
single-electron loss has been calculated especially for
hydrogen-like Pb$^{81+}$ and Au$^{78+}$ projectiles in collision
with neutral target atoms at collision energies of 160~GeV/u and
10.8~GeV/u, respectively~\cite{Matvgsi}. Based on these equations,
moreover, a method was developed for calculating the `energy loss'
of fast, heavy projectiles in collision with neutral
targets~\cite{Matv3}.

In principle, Eq.~(\ref{sigm}) could be applied together with the
probabilities in Eq.~(\ref{8m}) in order to compute the cross
sections for the loss of electrons from the projectile and for
both, a single or multiple ionization of electrons. In practice,
however, this is still hardly feasible for many-electron
projectiles since the summation over $[k]$ in Eq.~(\ref{sigm})
cannot be carried out explicitly in this case. Nevertheless, both
equations (\ref{8m}) and (\ref{sigm}) can be utilized in order to
obtain an expression that relates the cross section for the loss
of $N \,<\,N_p$ electrons, $\sigma^{N+}$, to the cross section
$\sigma^{N_p+}$ for the simultaneous ionization of all electrons.
Together with the assumptions from above, it was shown especially
in Ref.~\cite{Matv2} that the probability $W^{N+}({\bf b})$ for
the loss of $N$ electrons at a given impact parameter ${\bf b}$
can be written in terms of the single-electron form factors
$p_i({\bf b})$ as
\begin{eqnarray}
   W^{N+} ({\bf b}) & = & \frac{N_p}{(N_p-N)!N!}
   \nonumber \\[0.2cm]
   &  & \hspace*{0.1cm}
   \prod_{i=1}^{N_p-N} p_i({\bf b})
   \prod_{j=N_p-N+1}^{N_p}(1- p_j({\bf b})) ,
\end{eqnarray}
and where these form factors are given by
\begin{eqnarray}
\label{prob1}
   p_i({\bf b}) & = &
   \nonumber \\[0.2cm]
   &  & \hspace*{-1.5cm}
   \int d^3 k_i \, \left|\int d^3 r_i\Psi^{*}_{{\bf k}_i}({\bf r}_i)
   \exp\{-i\chi_i({\bf b},{\bf r}_i)\}\phi_i({\bf r}_i) \right|^2  .
\end{eqnarray}
Following Ref.~\cite{Matv2}, we next replace the single-electron form
factors by those as obtained from the average over all electrons with
orbital angular momentum $l$ and magnetic projection $m$ of the given shell
with principal quantum number $n$,
\begin{widetext}
\begin{eqnarray}
\label{eqaje21}
   p(b) & = & \frac{1}{n_0} \: \sum_{n=1}^{n_0}\:
              \frac{1}{n^2} \: \sum_{l,m} \:
          \int d^3k  \left|
          \int d^3r \, \Psi^{*}_{{\bf k}}
          \exp\{-i\,\chi({\bf b},{\bf r}) \} \phi_{nlm}({\bf r})
          \right|^2 \;,
\end{eqnarray}
and which has the advantage to depend only on the modulus $b$ of the
impact parameter. By taking this average, the probability $W^{N+}({\bf b})$
for the loss of $N$ electrons can then be expressed as
\begin{equation}
\label{prob}
   W^{N+}(b) \:=\: \frac{N_p}{(N_p-N)!N!}\sum_{m=0}^N
                   (-1)^m\frac{N!}{(N-m)!m!} \: \{ p(b)\}^{N_p-N+m} \, ,
\end{equation}
while the integration over the impact parameter $b$ gives finally
rise to
\begin{eqnarray}
\label{eqaf1}
   \sigma^{N+} & = &
   \frac{N_p !\,\sigma^{N_p+}}{(N_p-N)!\,N!}
   \sum_{m=0}^{N_p- N}(-1)^{m}
   \left(\frac{Z_p+N_p}{Z_p +N+m}\right)^{2} \:
   \frac{(N_p-N)!}{(N_p-N-m)!\:m!} \:
   \left(\frac{N_p}{N+m}\right)^{\kappa} \{p\,(b_{0},\,E)\}^{\,N- N_p+m} .
\end{eqnarray}
\end{widetext}
In this cross section expression for the loss of $N$ electrons,
$b_0$ hereby refers to the value of the impact parameter for which
the inelastic form factor takes its maximum, and $\kappa$ is a
quantity that characterizes the behavior of the function $p(b)$
near to this maximum, while $\Psi_{\bf k}$ is the  final-state
wave function of the projectile electron.

In general, neither the total cross section $\sigma^{N_p+}$ for the loss
of all electrons, nor the characteristic exponent $\kappa$, nor the
inelastic form factor $p(b)$ are known with sufficient accuracy from
(ab-initio) theory in order to make direct use of Eq.~(\ref{eqaf1}) and
to calculate from it the cross sections for the loss of $N$ electrons.
However, since for any given collision energy $E$, all the cross sections
$\sigma^{N+}\;\, (N=1,\,...,\,N_p)$ only depend on these three
quantities, Eq.~(\ref{eqaf1}) can be utilizes together with three
experimentally known cross sections in order to determine (numerically)
the values of $\sigma^{N_p+},\,\kappa$ and $p(b_0,\,E)$ within the
framework as outlined above. From these values, then, all the other
cross sections can be easily determined by applying Eq.~(\ref{eqaf1}).
While, for such a set of non-linear equations (\ref{eqaf1}), formally
of course quite different solutions may exist for $\sigma^{N_p+},\:\kappa$
and $\,p(b_0,\,E)\,$, all these parameters should be real and positive for
physical reasons. To our experience and up to the present, there has been
found only \textit{one} solution which fulfills this requirement.

More often than not, three cross sections $\sigma^{N_1+},\,
\sigma^{N_2+}$ and $\sigma^{N_3+}$ are either know or can be
measured for a given collision system. For the collision of
$U^{28+}$ projectiles with neutral argon (rest gas) atoms, for
example, the cross section for the loss of 1, 2, and up to 15
electrons have been measured in the Ref.~\cite{Olson}, and were
compared with the results of LOSS code calculations and the
classical trajectory Monte Carlo method. By applying the method
above, moreover, further cross sections for the multiple loss of
electrons have been calculated also for $U^{28+}$ and $U^{10+}$
projectiles~\cite{Matv2}. Here, we shall not follow these prior
computational lines further but extend the method in order to
apply it for different collision energies and/or targets.

To do so, let us note first that the cross section for the loss of
$N_p$ electrons is for $N_p \gg 1$ proportional to the $N_p$-th
power of the one-electron inelastic form factor, $\sigma^{N_p+}
\,\sim\, p(b,\,E)^{\,N_p}$~\cite{Matv2}. Therefore, the ratio
\begin{eqnarray}
\label{eqaf2}
   \frac{\sigma^{N_p+}}{\left[p(b_{0})\right]^{\,N_p}} & = & a
\end{eqnarray}
is found to be independent not only from the energy and the
initial charge state of the projectile but also with regard to the type of
the target atoms. To make further use of this observation, let us
introduce the term `collision system' in order to denote a
particular reaction scenario with given (type and charge state of the)
projectile and target atoms as well as with given collision energy.
If, for such a system, the cross sections are known for the loss of
\textit{three} different numbers of electrons, Eq.~(\ref{eqaf1})
can be utilized (as said before) to determine the cross sections
also for any other number of electrons. In addition, we can use
the cross section of some collision system A to determine the cross
section of any other collision system B, provided that \textit{two} cross
sections are known already for this system. The third cross
section value for system B is then simply obtained from the
relation~(\ref{eqaf2}) or, equivalently, from
\begin{eqnarray}
\label{eqaf3}
   \frac{\sigma^{N_p+}_{A}}{\left[p_{A}\right]^{N_p}} & = &
   \frac{\sigma^{N_p+}_{B}}{\left[p_{B}\right]^{N_p}} \, .
\end{eqnarray}
In Section~III, we shall apply this recipe to derive the cross
section for the loss of $N$ electrons for systems for which only
two cross sections are (supposed to be) known from experiment. The
validity of this approach can be then tested easily by comparing
the cross section with the data as obtained from the prior
knowledge of `three' cross sections for the system. Again, the
determination of the three parameters $\sigma^{N_p+},\,\kappa$ and
$p(b_0,\,E)$ then follows lines similar as discussed above and by
making use of the relation (\ref{eqaf2}) as the additional (third)
equation.

\begin{figure}[t]
\begin{center}
\includegraphics[width=5.7cm,angle=-90]{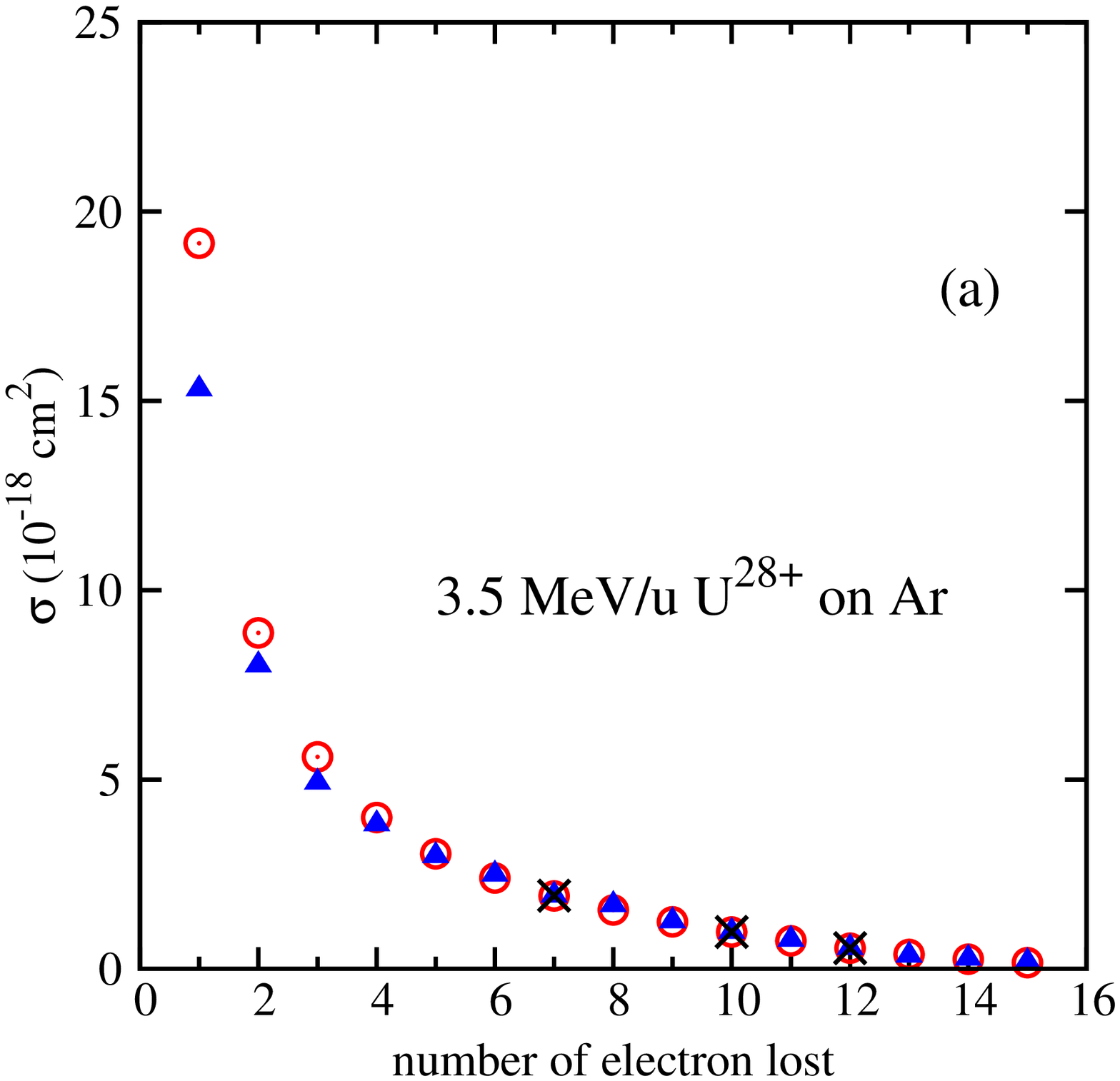}
\end{center}
\begin{center}
\includegraphics[width=5.7cm,angle=-90]{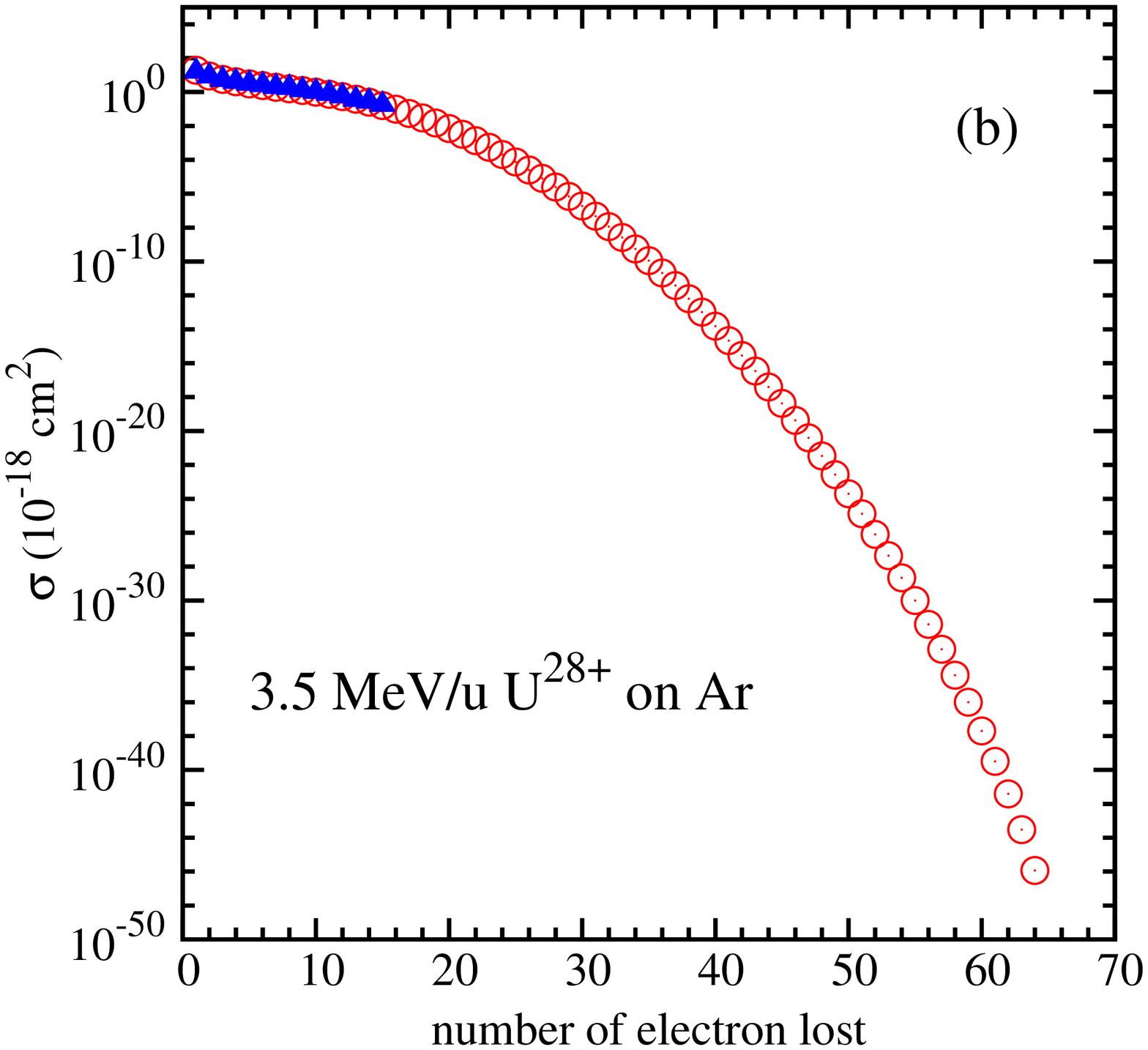}
\end{center}
\caption{(Color online): Cross sections (in $10^{-18}~cm^2$) for a
multiple loss of projectile electrons in collisions of U$^{28+}$
ion with neutral Ar atoms at collision energy E = 3.5~MeV/u, and
as function of the number $N$ of lost electrons; (a) for $N =
1,...,15$ and (b) for $N = 1,...,64$. Computations from this work
(open circles) are compared with the experimental data (filled
triangles) from Ref.~\cite{Olson}. The crosses indicate the
experimental data that were utilized for the computations.}
\label{fig1}
\end{figure}

Finally, if two collision systems A and B differ only by the
(collision) energy $E_A \ne E_B$, we have the additional condition
$\kappa_A =\kappa_B$, and the prescription above simplifies
considerably. In this case, only \textit{one} experimental cross
section need to be known for the energy $E_B$, while the other two
cross sections (parameters in Eq.~(\ref{eqaf1})) can be utilized
from the collision of the projectile and target at the
energy $E_A$. In the next section, results will be shown and
discussed especially for the multiple electron loss of xenon and
uranium ions in collision with different target materials. These
ions are important for the design of the FAIR facility in
Darmstadt, and most experimental data are available for them.

\begin{figure}[t]
\begin{center}
\includegraphics[width=5.7cm, angle=-90]{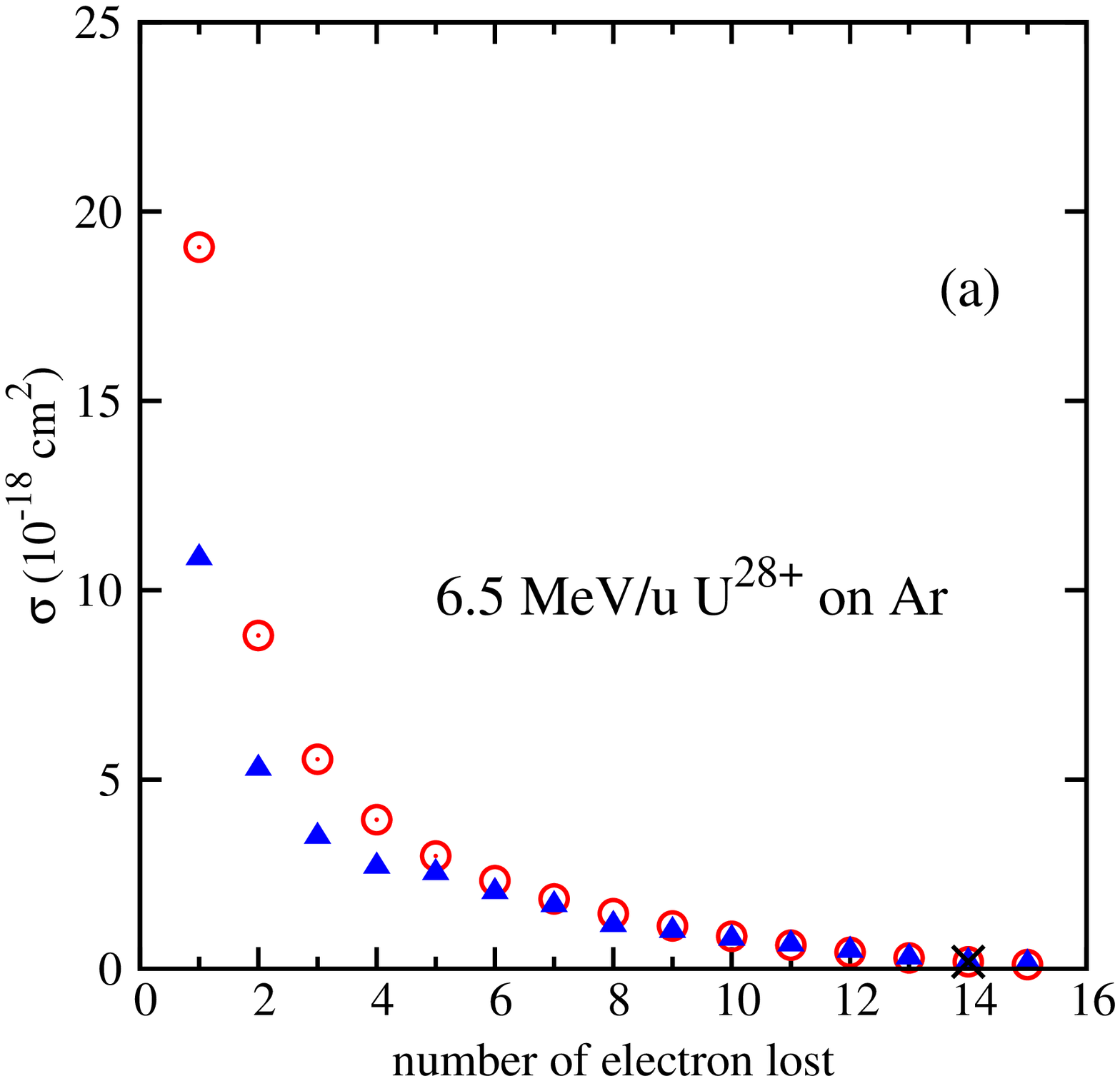}
\end{center}
\begin{center}
\includegraphics[width=5.7cm, angle=-90]{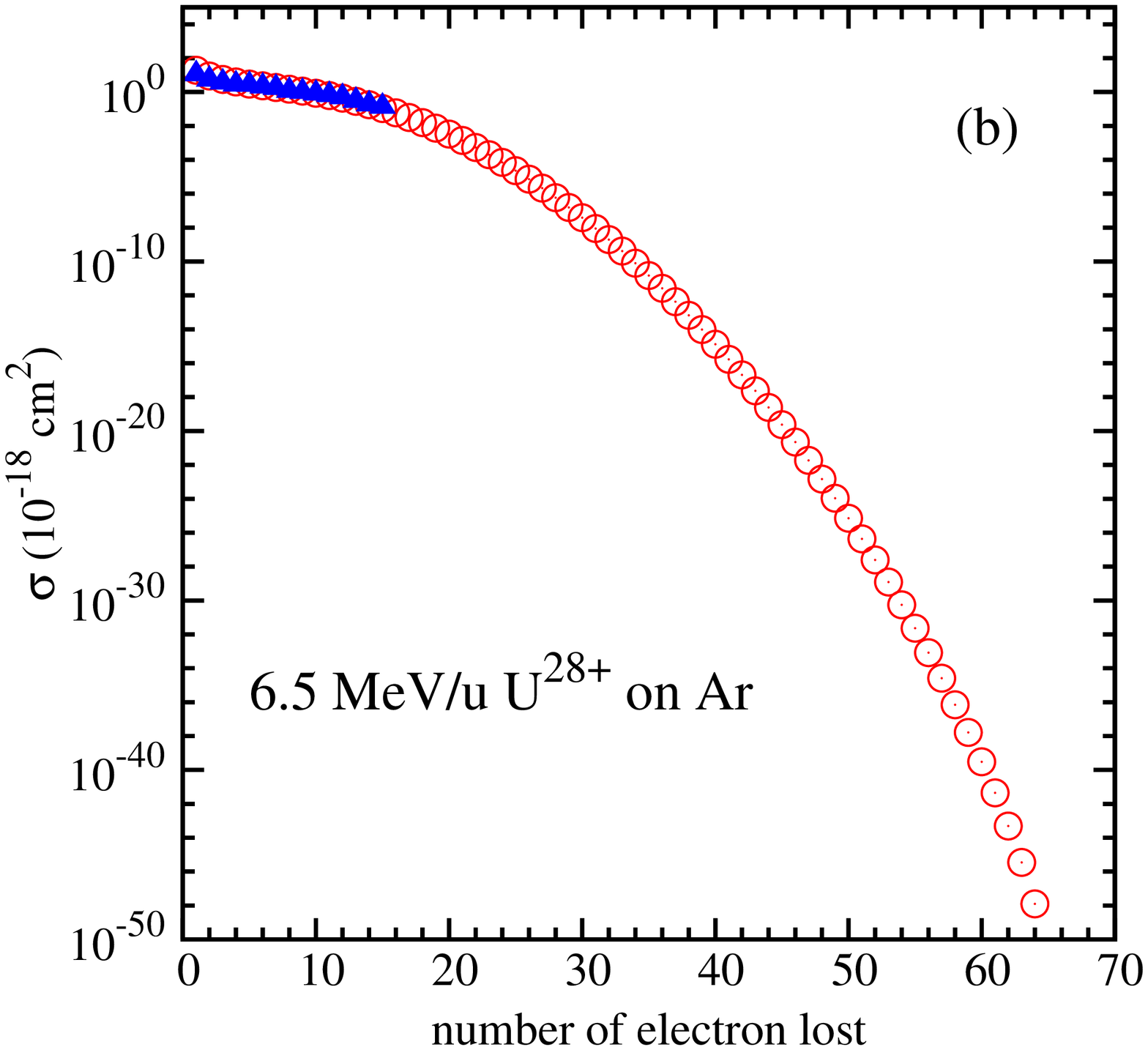}
\end{center}
\caption{(Color online): The same as Fig.~1 but for the collision
energy E = 6.5~MeV/u. For the theoretical cross sections, only the
single experimental value for the loss of $N=14$ electrons has
been utilized at this energy; see text for discussion.}
\label{fig2}
\end{figure}

\section{Results and Discussion}

Cross sections for the electron loss of fast, heavy projectiles
have been measured for a number of collision systems. For example,
Olson \etal{}~\cite{Olson} display measured cross sections for the
electron loss of up to 15 electrons for collisions of U$^{28+}$
ions with neutral Ar atoms. These cross sections are shown in
Fig.~1(a) for the projectile energy E = 3.5~MeV/u and in Fig.~2(a)
for 6.5~MeV/u, respectively, and are compared with our theoretical
prediction following the prescription above. To generate these
theoretical values, the experimental cross sections for the
(simultaneous) loss of 7, 10, and 12 electrons have been utilized
[cf.~crosses in Figs.~1(a)]. Theoretical data are shown for the
loss of $N = 1,...,15$ electrons in Fig.~1(a) and for $N =
1,...,64$ in Fig.~1(b). Very similar results within a few percent
were obtained if just three other experimental cross sections are
applied. In general, however, the cross sections for a loss of
$N<5$ electrons should not be used in order to obtain the cross
sections for large $N$ because they often appear more sensitive to
many-electron effects that are not incorporated into the model.
Fig.~2 displays analogue cross sections for U$^{28+}
\,\rightarrow\:$Ar collisions but for the collision energy E =
6.5~MeV/u. For the theoretical cross sections in Fig.~2, only the
(single) cross section for the loss of $N=14$ electrons have been
utilized from the experiments at 6.5~MeV/u, while the other
information were obtained by using the ratio~(\ref{eqaf2}) and
$\kappa$ from the data for E =3.5~MeV/u in Fig.~1. Typically, good
agreement between theory and experiment is found, and only the
cross sections for the loss of just a very few electrons $(N \,=\,
1,...,4)$ are slightly overestimated by our model. As expected,
the electron-loss cross sections for the collision energy
6.5~MeV/u are larger than for 3.5~MeV/u, independent of how many
electrons are lost from the projectile.

\begin{figure}[t]
\begin{center}
\includegraphics[width=5.7cm, angle=-90]{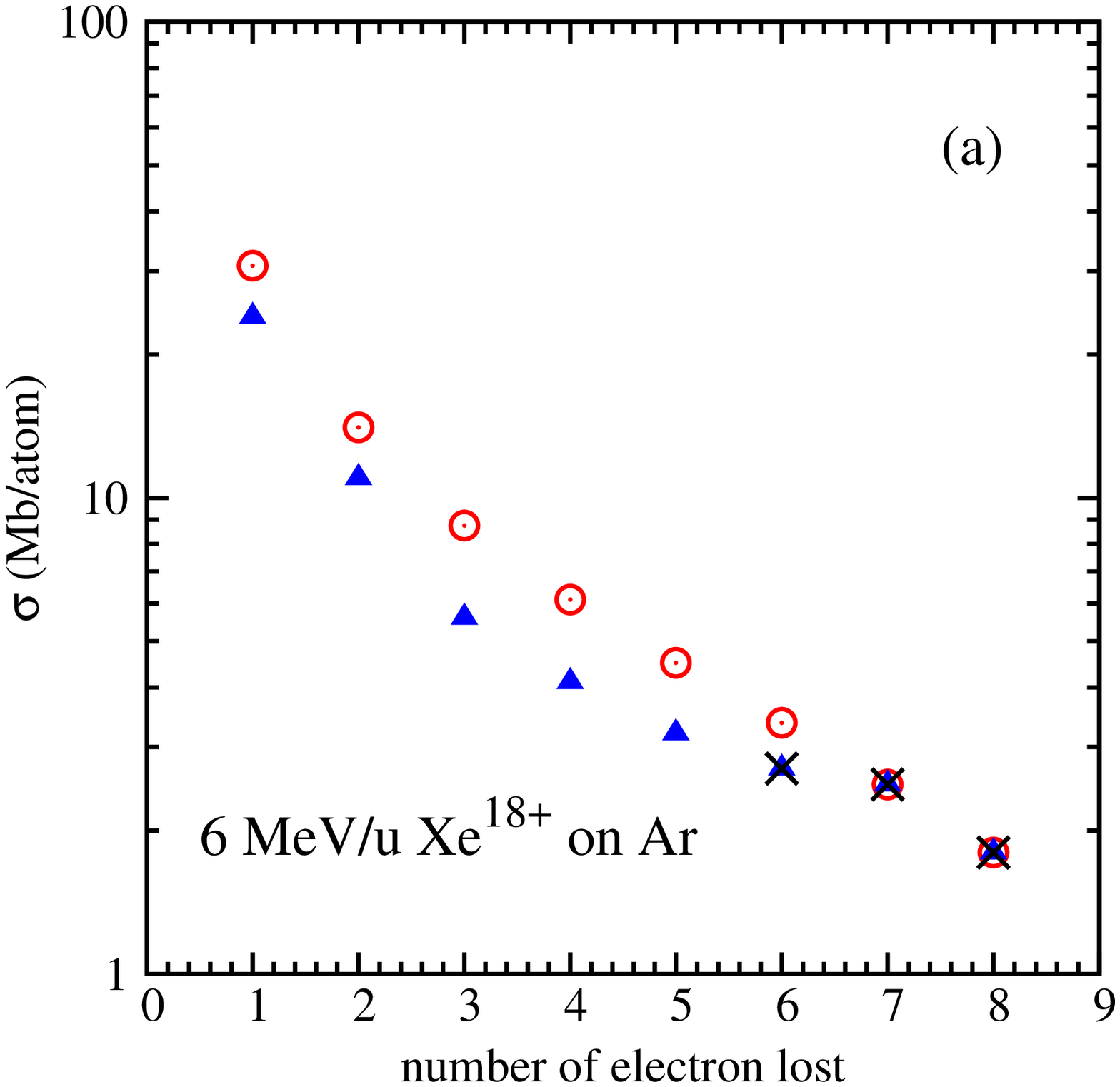}
\end{center}
\begin{center}
\includegraphics[width=5.7cm, angle=-90]{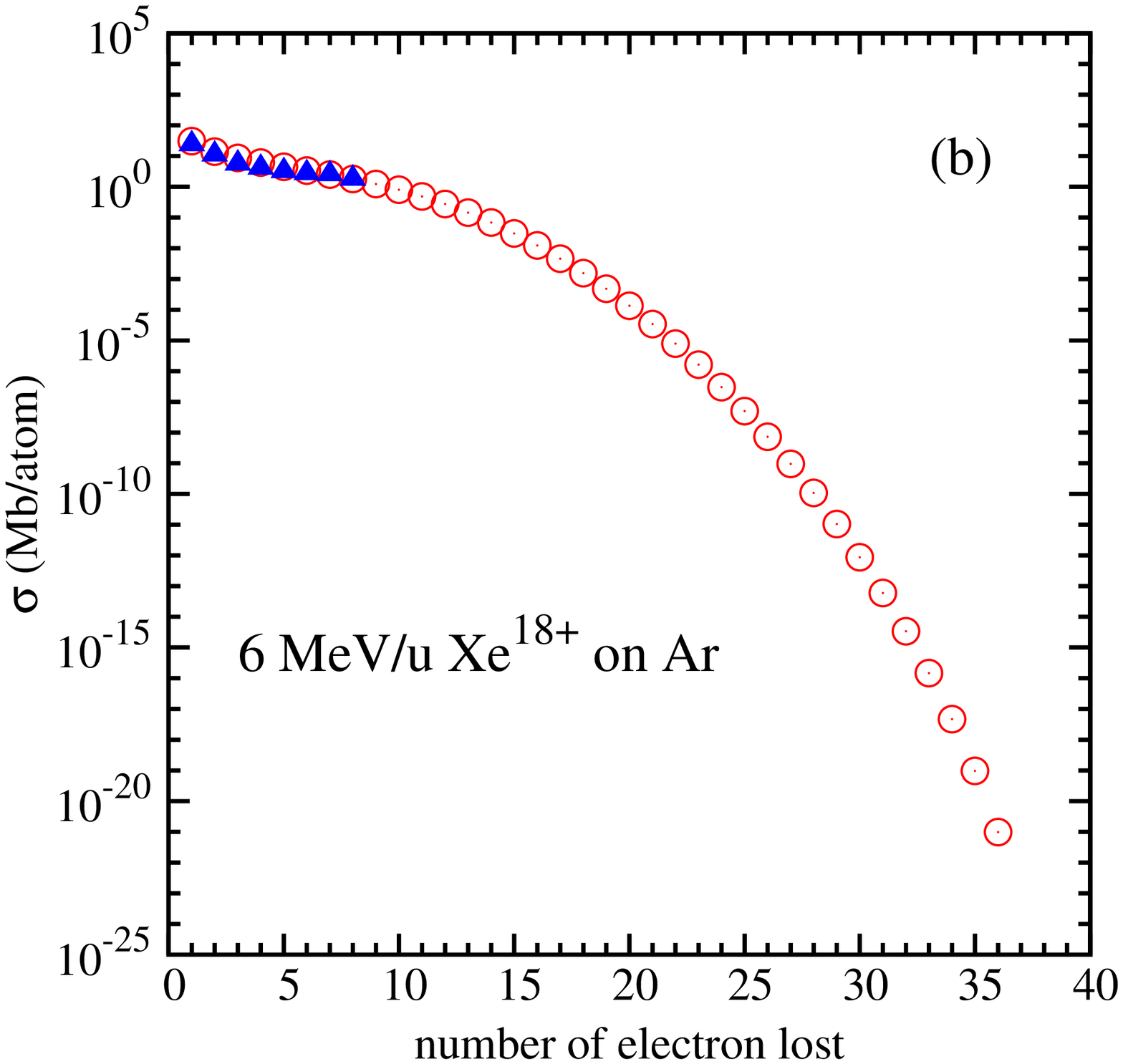}
\end{center}
\caption{(Color online): The same as in the Figs.~1 but for the
collision systems Xe$^{18+} \rightarrow$ Ar and the energy E =
6~MeV/u; (a) for $N = 1,...,8$ and (b) for $N = 1,...,36$.}
\label{fig3}
\end{figure}

\begin{figure}[t]
\begin{center}
\includegraphics[width=5.7cm, angle=-90]{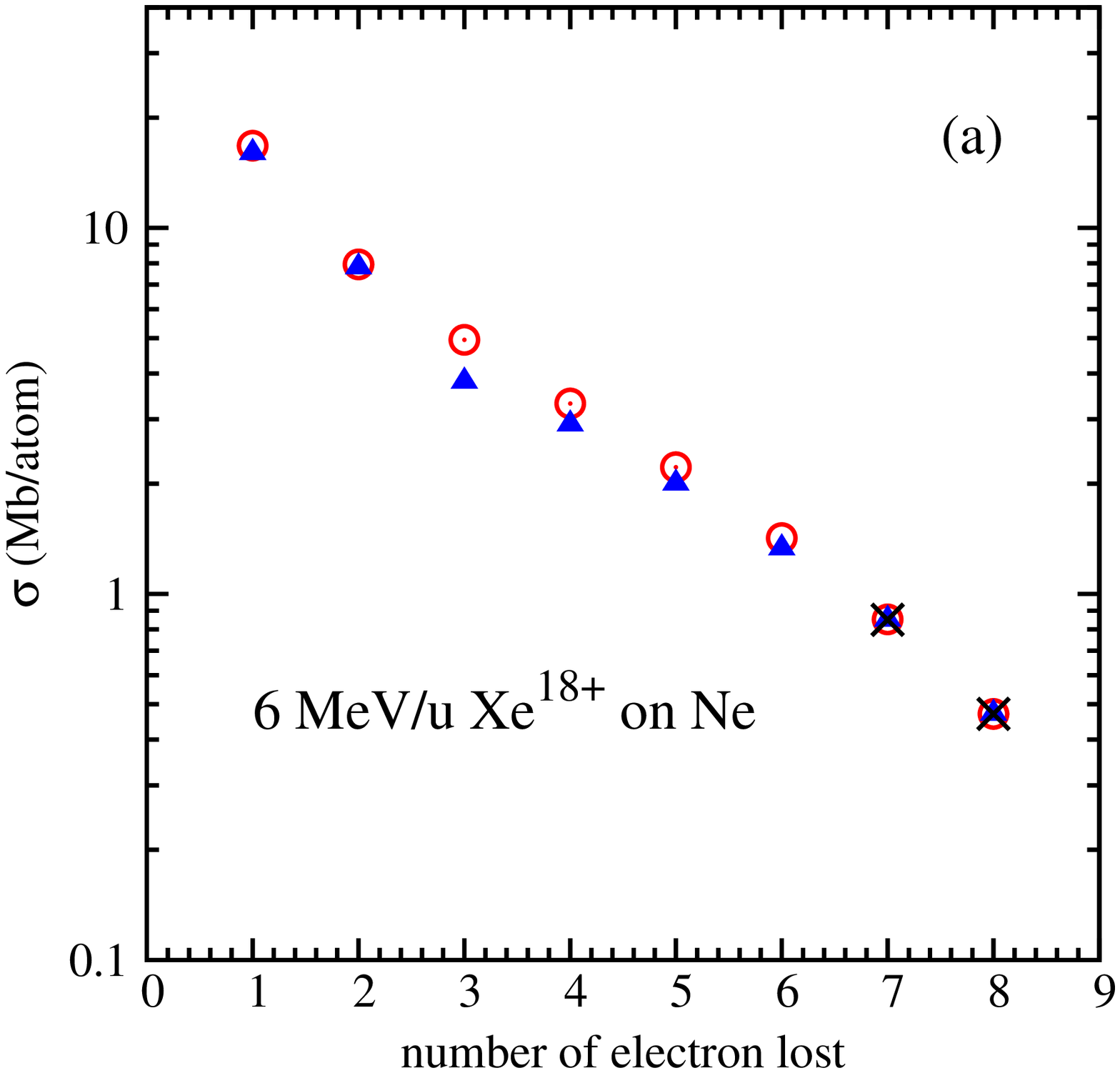}
\end{center}
\begin{center}
\includegraphics[width=5.7cm, angle=-90]{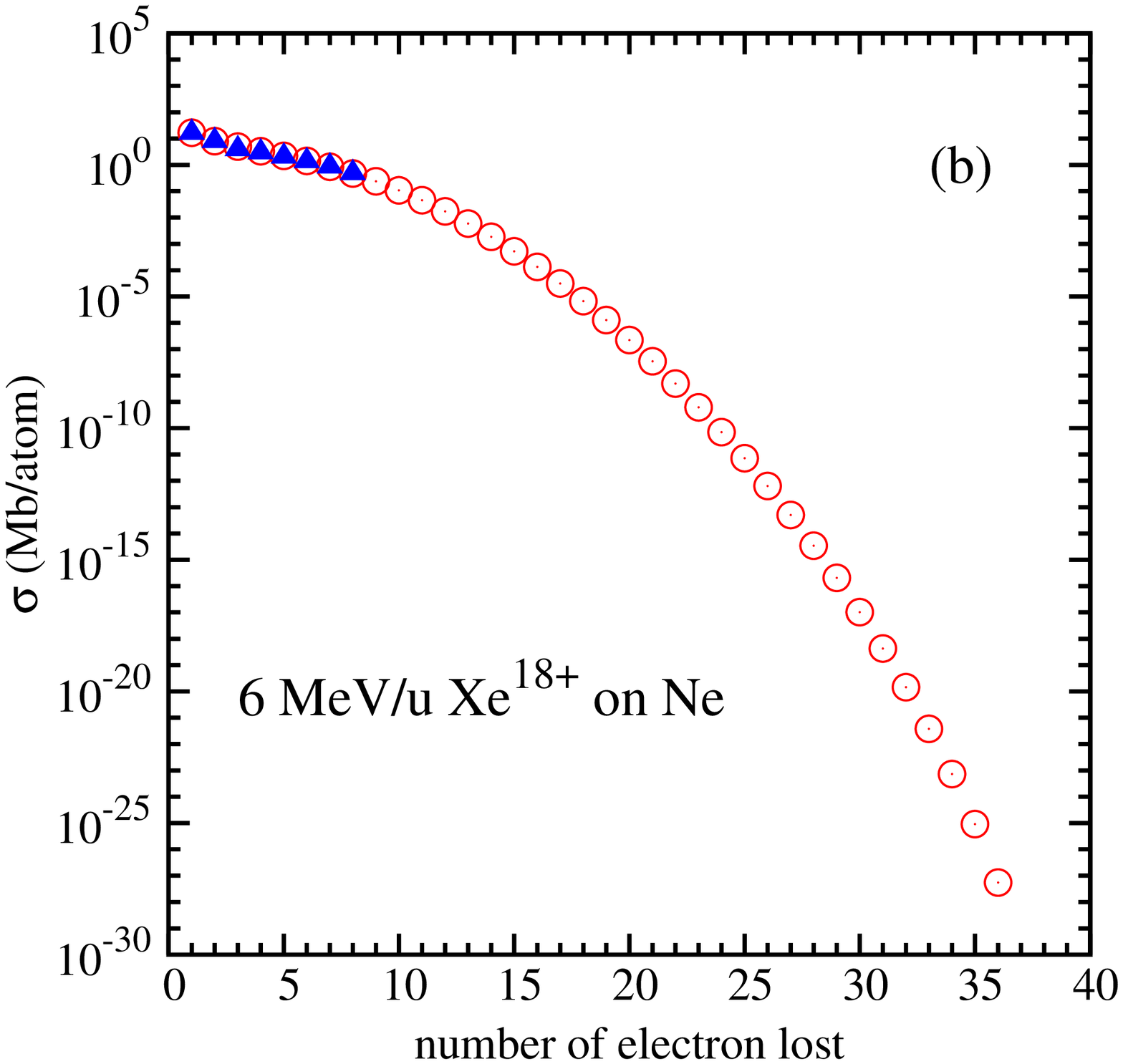}
\end{center}
\caption{(Color online): The same as in the Fig.~3 but for the
collision systems Xe$^{18+} \rightarrow$ Ne and the energy E =
6~MeV/u; (a) for $N = 1,...,8$ and (b) for $N = 1,...,36$.}
\label{fig4}
\end{figure}

\begin{figure}[t]
\begin{center}
\includegraphics[width=5.7cm, angle=-90]{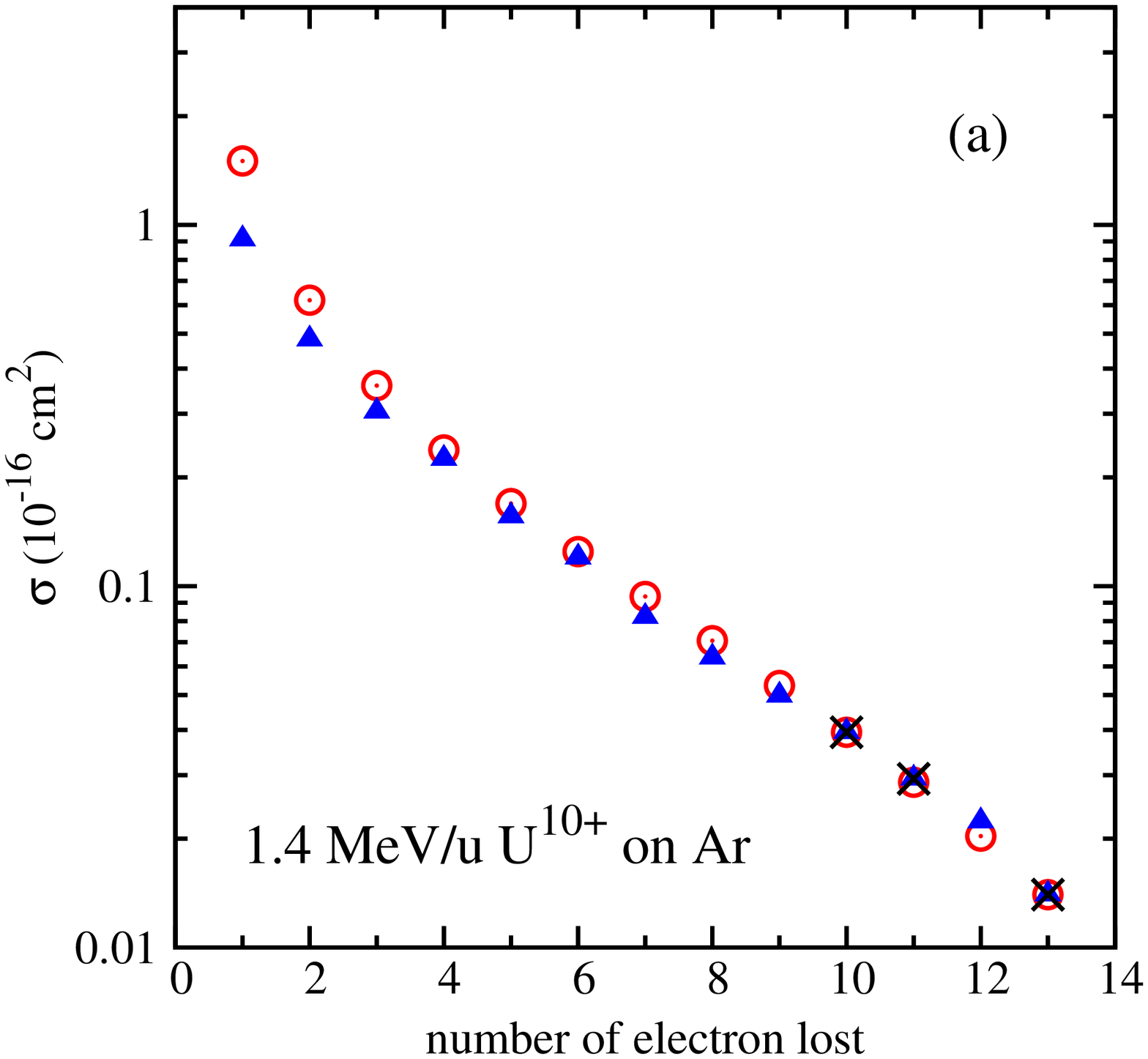}
\end{center}
\begin{center}
\includegraphics[width=5.7cm, angle=-90]{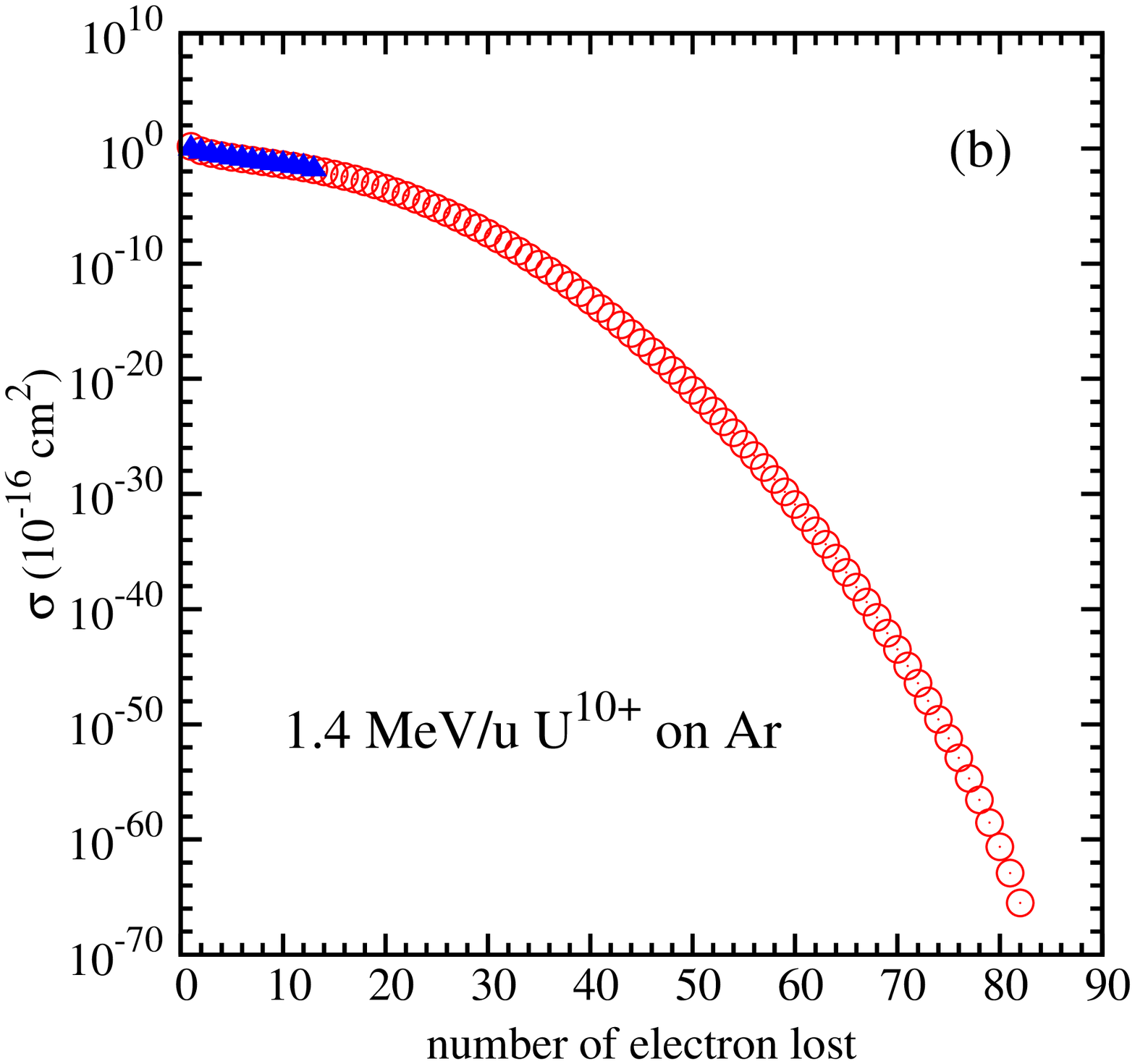}
\end{center}
\caption{(Color online): The same as in the Fig.~3 but for the
collision systems U$^{10+} \rightarrow$ Ar and the collision
energy E = 1.4~MeV/u; (a) for $N = 1,...,13$ and (b) for $N =
1,...,82$. Theoretical cross sections are compared with the
measurements by DuBois \etal{}~\cite{DuBois} as far as available.}
\label{fig5}
\end{figure}

\begin{figure}[t]
\begin{center}
\includegraphics[width=5.7cm, angle=-90]{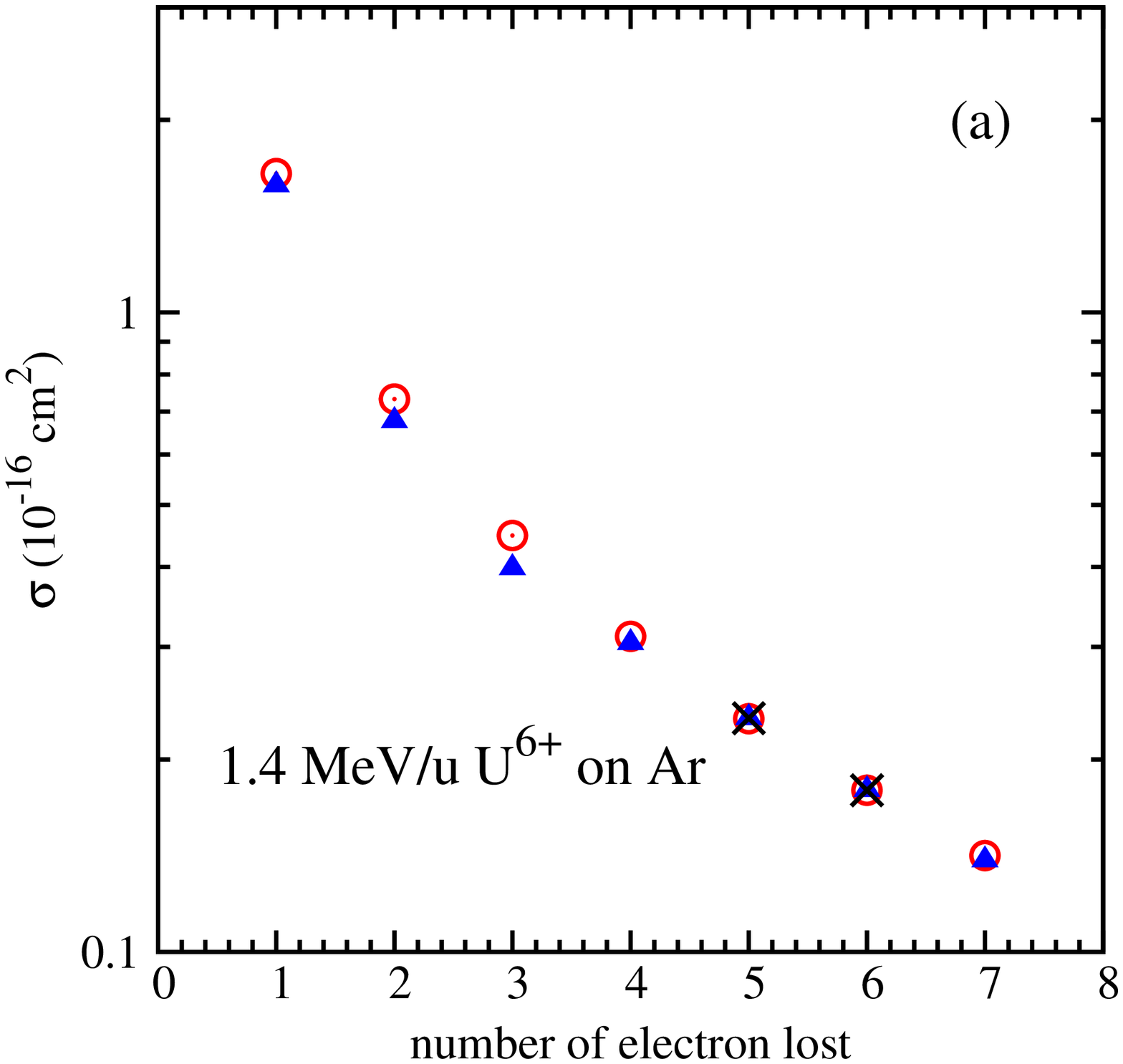}
\end{center}
\begin{center}
\includegraphics[width=5.7cm, angle=-90]{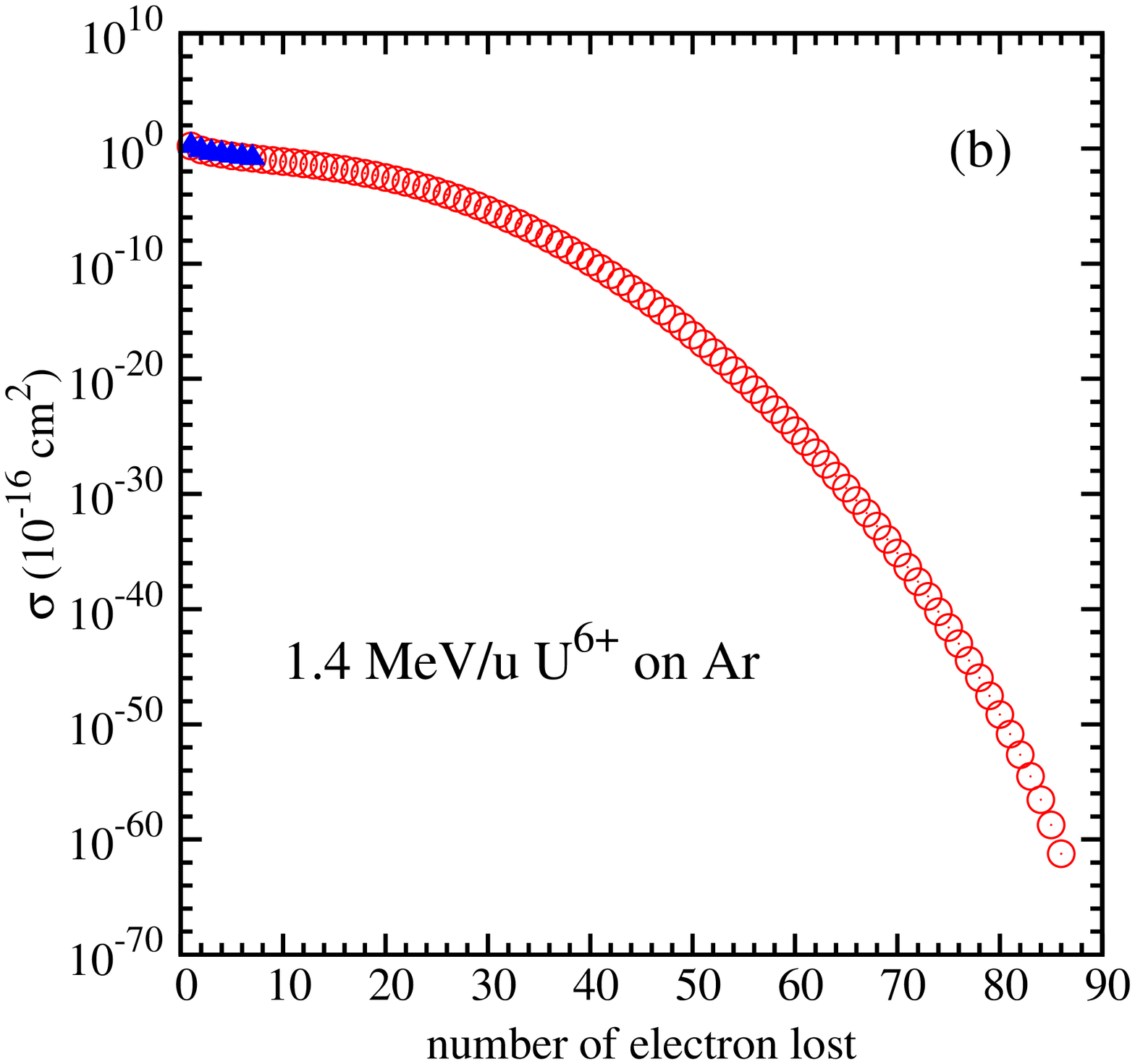}
\end{center}
\caption{(Color online): The same as in the Fig.~5 but for the
collision systems U$^{ 6+} \rightarrow$ Ar  and the collision
energy E = 1.4~MeV/u; (a) for $N = 1,...,7$ and (b) for $N =
1,...,86$. Theoretical cross sections are compared with the
measurements by DuBois \etal{}~\cite{DuBois} as far as available.}
\label{fig6}
\end{figure}

\begin{figure}[t]
\begin{center}
\includegraphics[width=5.7cm,angle=-90]{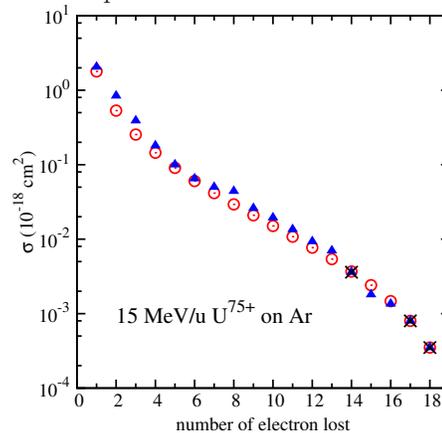}
\end{center}
\caption{(Color online): Cross sections (in $10^{-18}~cm^2$) for a
multiple loss of \textit{target} electrons in U$^{28+}
\rightarrow$ Ar collisions at E = 15~MeV/u, and as function of the
number $N$ of lost electrons. The notation is the same as in
Fig.~1. Computations from this work are compared again with the
experimental data from Ref.~\cite{Olson}.} \label{fig7}
\end{figure}

Our model from above works well also if we consider two collision
systems that differ in the target atom. This is seen from
Figs.~3-4 which display the projectile electron-loss cross
sections for the two systems Xe$^{18+} \rightarrow$ Ar (cf.\
Figs.~3(a) and 3(b)) and Xe$^{18+} \rightarrow$ Ne (cf.\
Figs.~4(a) and 4(b)) at the collision energy E = 6~MeV/u. In
Fig.~3 (a,b), again, the theoretical cross sections are based on
three measured data by Watson \etal{}~\cite{Watson} for the loss
of 6, 7, and 8 electrons from the Xe$^{18+}$ projectile, while the
ratio~(\ref{eqaf2}) and only two cross sections (for the loss of 7
and 8 electrons) are used in Figs.~4(a,b). The theoretical cross
sections for Xe$^{18+} \rightarrow$ Ar collisions are slightly
overestimated  for the loss of just a few electrons but are in
excellent agreement with experiment for the Xe$^{18+} \rightarrow$
Ne collisions in Figs.~4(a,b). In Figs.~5 and 6, theoretical
results are shown for the collision of U$^{10+} \rightarrow$ Ar
and U$^{ 6+} \rightarrow$ Ar at the energy E = 1.4~MeV/u, i.e.\
for two different initial charges states of the projectile.
Experimental cross sections from  Ref.~\cite{DuBois} were utilized
in order to derive the theoretical data. Very good agreement
between theory and the experiments by DuBois \etal{}~\cite{DuBois}
is found in both cases.

So far, we have always considered the multiple loss of electrons
from the projectile ions in collision with different target atoms
and for various collision energies. As mentioned before, the same
model can be utilized also to predict the ionization of the
target. For this, only the nuclear charge $Z_p \leftrightarrow
Z_a$ and the (initial) number of bound electrons $N_p
\leftrightarrow N_a$ need to be interchanged in Eq.~(\ref{eqaf2}),
while the rest of the computational procedure remains rather
unchanged. In Fig.~7, our theoretical prescription is applied to
predict the cross sections for the multiple loss of target
electrons from argon in U$^{28+} \rightarrow$ Ar collisions at E =
15~MeV/u. Cross sections are displayed for single ionization and
up to the complete (18-fold) stripping of all electrons from Ar.
The theoretical data are compared with the experiments by Olson
\etal{}~\cite{Olson} and are found in excellent agreement. This
confirms in a practical manner that our model can be applied for
the multiple loss of electrons from both, the projectile and
target.

\section{Conclusions}

The multiple electron loss of heavy projectiles in fast ion-atom
collisions has been investigated. Based on the sudden perturbation
approximation, a model is developed to estimate the cross
sections for a multiple loss of electrons from both, the projectile
and target atoms, and up to their complete ionization. In this model,
only three (measured) cross sections
are needed from experiment in order to predict the loss of any
number $N$ of electrons for a given collision system. Moreover,
the model can be applied to different projectiles and targets if
the cross sections for the loss of at least two different numbers of
electrons are known, while the third cross section value can be
taken from any other system. Only a single measured cross section
is needed, moreover, if (two) other values for the same
projectile-target collision system are known at some different
collision energy. By making use of this model, calculations have
been carried out especially for the multiple electron-loss cross
sections of U$^{28+} \rightarrow$ Ar, U$^{10+} \rightarrow$ Ar,
U$^{6+} \rightarrow$ Ar as well as for Xe$^{18+} \rightarrow$ Ar
and Xe$^{18+} \rightarrow$ Ne, and for different energies. The
model is simple and can be utilized without that large computer
resources are required.

%
%

\section*{ACKNOWLEDGMENTS}

This work has been supported by an INTAS grant from GSI
(Ref.~Nr.~8530) and by the grant of the Uzbek Academy of Sciences
(FA-F2-084).

%
%

\end{document}